\documentclass[twocolumn,secnumarabic,amssymb, nobibnotes, aps, prb, superscriptaddress]{revtex4-2}
\usepackage{lipsum}
\usepackage{titlesec}

\titlespacing\section{0pt}{12pt plus 4pt minus 4pt}{1pt plus 20pt minus 2pt}
\usepackage{xcolor}
\usepackage[colorlinks, allcolors=blue]{hyperref}
\usepackage{tabularx}
\usepackage{amsmath}
\usepackage{comment}
\usepackage{chemformula}
\usepackage{afterpage}
\usepackage{siunitx}
\usepackage{gensymb}
\usepackage{placeins}
\usepackage{graphicx}
\usepackage{float}
\usepackage{booktabs}
\usepackage{multirow}
\usepackage{array}
\usepackage{setspace}
\graphicspath{{Figs/}}
\usepackage{siunitx}
\usepackage{hhline}
\usepackage{xfrac}
\usepackage{float,graphicx}
\usepackage{mathtools}
\usepackage{listings}
\usepackage{amssymb}
\usepackage{titlesec}
\usepackage{amsfonts}

\usepackage[version=4]{mhchem}

\usepackage{epstopdf}
\catcode`@11
\def\seceqaa{\@addtoreset{equation}{section}
\def\theequation{A\arabic{equation}}}
\def\seceqbb{\@addtoreset{equation}{section}
\def\theequation{B\arabic{equation}}}
\def\seceqcc{\@addtoreset{equation}{section}
\def\theequation{C\arabic{equation}}}
\def\seceqdd{\@addtoreset{equation}{section}
\def\theequation{D\arabic{equation}}}
\def\seceqee{\@addtoreset{equation}{section}
\def\theequation{E\arabic{equation}}}
\def\seceqff{\@addtoreset{equation}{section}
\def\theequation{F\arabic{equation}}}
\def\seceqgg{\@addtoreset{equation}{section}
\def\theequation{G\arabic{equation}}}
\def\seceqhh{\@addtoreset{equation}{section}
\def\theequation{H\arabic{equation}}}
\catcode`@11

\begin{document}

\title{Role of antisite disorder in dictating the Ising anisotropy and ferromagnetic ground state of NdGaSi}

\author{Anyesh Saraswati}
\affiliation{S. N. Bose National Centre for Basic Sciences, Salt Lake City, Kolkata-700106, India}

\author{Sourav Kanthal}
\affiliation{S. N. Bose National Centre for Basic Sciences, Salt Lake City, Kolkata-700106, India}

\author{Subhajit Nandy}
\affiliation{Deutsches Elektronen-Synchrotron DESY, Notkestr. 85, Hamburg, 22607, Germany}

\author{Jhuma Sannigrahi}
\affiliation{School of Physical Sciences, Indian Institute of Technology Goa, Farmagudi, Goa 403401, India}

\author{Matthias Gutmann}
\affiliation{ISIS Neutron and Muon Source, Science and Technology Facilities Council, Rutherford Appleton Laboratory, Chilton, Didcot OX11 0QX, United Kingdom}

\author{Nitesh Kumar}\email{nitesh.kumar@bose.res.in}
\affiliation{S. N. Bose National Centre for Basic Sciences, Salt Lake City, Kolkata-700106, India}

\begin{abstract}

Single-ion anisotropy and exchange interactions together set the magnetic ground state of rare-earth intermetallics, yet the role of chemical disorder on the ligand sublattice remains little explored. Here, we show that Ga/Si antisite disorder in NdGaSi stabilises the centrosymmetric $I4_1/amd$ structure and raises the Nd$^{3+}$ site symmetry to $D_{2d}$, which forbids all $m=\pm2$ Stevens operators in the crystalline electric field (CEF) Hamiltonian. Magnetization, specific heat, single-crystal neutron diffraction, and x-ray absorption spectroscopy establish the disordered structure and a collinear ferromagnetic order below $T_\mathrm{C}\approx11$~K with Ising anisotropy along $c$. A CEF analysis that fits the susceptibilities along both axes and the magnetic specific heat simultaneously yields a ground-state doublet that is $\sim97\%$ pure, in contrast to the strongly mixed ground state of the Al/Si-ordered NdAlSi ($C_{2v}$ site symmetry of Nd$^{3+}$). We argue that this ground-state purity makes the ordered moment rigid against the modulation due to frustrated exchange, selecting a $\mathbf{k}=0$ ferromagnet in NdGaSi instead of the incommensurate ferrimagnet of NdAlSi. Antisite disorder, thus, acts as a handle on CEF ground-state purity and can, thereby, select between fundamentally different magnetic ground states.

\vspace{3mm}
\end{abstract}

\maketitle
\section{INTRODUCTION}
\vspace{3mm}

The interplay between magnetism and topological electronic states has become a central focus in condensed matter physics, giving rise to phenomena such as the anomalous Hall effect (AHE), chiral magnetic textures, and Weyl fermions \cite{armitage2018weyl,yan2017topological,kumar2020topological,nagaosa2010anomalous,liu2018giant,nagaosa2013topological,tokura2020magnetic,kurumaji2019skyrmion,hasan2021weyl}. In rare-earth intermetallics, the partially filled 4\textit{f} shells provide large magnetic moments and strong spin-orbit coupling (SOC), while the crystalline electric field (CEF) lifts the degeneracy of the Hund's ground state (GS) multiplet, producing single-ion anisotropy that competes with inter-ionic exchange interactions \cite{skomski2009anisotropy,elliott2013magnetic,jensen1991rare,irkhin1988electron,johansson2007theory,bouaziz2024origin}. When combined with a noncentrosymmetric crystal structure, these ingredients can stabilise Weyl nodes, directly coupling magnetism to topology, as in the rare-earth based 4\textit{f} intermetallic family \textit{R}Al\textit{X} (\textit{R} = Ce, Pr, Nd, Sm, Gd; \textit{X} = Si, Ge) \cite{chang2018magnetic,yang2021noncollinear,alam2023sign,yao2023large,gaudet2021weyl,nag2024gdalsi,puphal2020topological,piva2023topological,meng2019large,dhital2023NAG,yang2023stripe}. These systems generally crystallize in the noncentrosymmetric tetragonal LaPtSi-type structure (space group $I4_1md$) \cite{kneidinger2013synthesis}, as confirmed by neutron diffraction and second harmonic generation measurements \cite{yang2021noncollinear,shoriki2024large,yang2020transition,gaudet2021weyl,yao2023large,nag2024gdalsi,nakano2026perfectly}.

\vspace{2mm}

The \textit{R}AlSi compounds show remarkably diverse magnetism, set by competing Ruderman--Kittel--Kasuya--Yosida (RKKY), superexchange (SX), and Dzyaloshinskii--Moriya (DMI) interactions together with single-ion anisotropy \cite{bouaziz2024origin}. CeAlSi is ferromagnetic (FM) below 8.2~K with a noncollinear texture \cite{yang2021noncollinear,alam2023sign}, and PrAlSi is FM below 17.8~K with two reentrant, glass-like transitions \cite{lyu2020nonsaturating}. NdAlSi instead orders as an incommensurate (INC) ferrimagnet (FIM) below 7.2~K that locks into a non-coplanar commensurate FIM below 3.3~K \cite{gaudet2021weyl,wang2022ndalsi}; an INC fan-type structure has also been proposed \cite{kurumaji2025note}. SmAlSi hosts a spiral order with a large topological Hall effect \cite{yao2023large,thaivalappil2026theory}, and GdAlSi shows two antiferromagnetic (AFM) transitions near 32~K and several field-induced metamagnetic (MMT) phases \cite{li2026field,nag2024gdalsi}. Bouaziz \textit{et al.} showed that frustrated short-range SX via \textit{p}--\textit{d} hybridization, rather than RKKY, drives the helical order in NdAlSi and SmAlSi: competition between intra- and interlayer SX gives a finite-$\mathbf{q}$ instability, while the CEF anisotropy decides between collinear FM (PrAlSi) and helical order (SmAlSi) \cite{bouaziz2024origin,yao2023large,lyu2020nonsaturating}.  In most studies, attention has focused on the rare-earth ion itself, with the non-magnetic sublattice taken as a fixed, ordered framework. Site mixing on such non-magnetic sublattices is, however,  common in rare-earth intermetallics and is generally reported as a structural detail, yet its effect on the CEF, and thereby on the magnetism, is rarely considered.

\vspace{2mm}

The Ga-based analogues \textit{R}Ga\textit{X} are much less understood. Single-crystal x-ray diffraction (XRD) shows that complete Ga/Si site mixing in silicide variants gives them the centrosymmetric $\alpha$-ThSi$_2$-type structure ($I4_1/amd$), raising the rare-earth site symmetry to $D_{2d}$, compared with $C_{2v}$ in ordered\textit{ R}AlSi systems \cite{darone2013rare}. Their magnetism differs markedly from the Al compounds. CeGaGe and PrGaGe are FM below 6.0 and 19.4~K, respectively, with strong uniaxial anisotropy and additional low-temperature anomalies \cite{ram2023magnetic}, and NdGaGe orders below 8.8~K with a large intrinsic anomalous Hall conductivity (AHC) \cite{ram2024crystalline,cao2025canting}. CeGaSi is AFM below 12~K with a weak in-plane FM component and a skew-scattering AHE \cite{gong2024anomalous,zhang2024magnetism,swami2025crystalline}, PrGaSi is FM below 23~K \cite{swami2025crystallinepgs}, SmGaSi is AFM ($T_\mathrm{N}=11$~K) with weak antilocalisation \cite{saraswati2026single}, and GdGaSi is AFM below 19~K with a spin-reorientation and positive magnetoresistance \cite{panda2026investigation}. NdGaSi stands out: it is FM below 11~K with the easy axis along $c$ and an AHC of $\approx1730~\Omega^{-1}$cm$^{-1}$ \cite{saraswati2026accessing,zhong2025large}, whereas NdAlSi is an INC FIM with no measurable AHE \cite{gaudet2021weyl,wang2022ndalsi,zhang2024abnormally,li2023emergence}. However, the only structural difference between these two compounds is the way the non-magnetic isoelectronic atoms  (Al and Ga) position themselves in the structure. Yet the two compounds are closely related: Al and Ga are isoelectronic, Nd is trivalent in both, and the main structural difference is whether the non-magnetic atoms are ordered (NdAlSi) or randomly mixed (NdGaSi). The pair therefore offers a clean test of whether disorder on the non-magnetic sublattice can select the magnetic GS, with direct consequences for the anomalous Hall response. 

\vspace{2mm}

To understand the effect of Ga/Si antisite disorder and the consequent change in Nd site symmetry, we carry out a detailed CEF analysis of NdGaSi and compare it with the Al counterpart NdAlSi. Single-crystal XRD of NdGaSi \cite{saraswati2026accessing,darone2013rare} places Nd$^{3+}$ at a $D_{2d}$ site of the $I4_1/amd$ structure, and Nd $L_3$-edge x-ray absorption spectroscopy (XAS) confirms this picture at the local level: both compounds contain trivalent Nd, but the extended x-ray absorption fine structure (EXAFS) resolves distinct Nd--Al and Nd--Si distances in NdAlSi, whereas NdGaSi shows a single, narrow distribution of Nd--Ga/Si distances broadened by static Ga/Si disorder. Single-crystal neutron diffraction (SCND) reveals a collinear FM order along $c$ with $\mathbf{k}=(0,0,0)$. Fitting the susceptibility along both crystallographic axes and the magnetic specific heat simultaneously yields the CEF parameters, level scheme, and GS wavefunction. The $D_{2d}$ symmetry forbids all $m=\pm2$ terms ($B_2^2$, $B_4^2$, $B_6^2$) that are allowed at the $C_{2v}$ site of NdAlSi, so the NdGaSi GS is a nearly pure ($\approx97\%$) $|\pm7/2\rangle$ Kramers doublet, in contrast to the strongly mixed GS of NdAlSi ($\approx76\%$ $|\pm9/2\rangle$) \cite{yang2025crystalline}. We argue that this rigidity of the Ising doublet results in a collinear FM GS in NdGaSi over the frustrated INC FIM of NdAlSi. The change of the average Nd site symmetry from $C_{2v}$ to $D_{2d}$, caused by Ga/Si disorder on the non-magnetic sublattice, thus leads to a pure CEF GS and a more stable magnetic order than in the ordered counterpart. This suggests that ligand order, usually regarded as a structural detail, can influence the choice between distinct magnetic GS in rare-earth intermetallics.

\section{EXPERIMENTAL METHODS}
\vspace{2mm}
Single crystals of NdGaSi were synthesized using a gallium self-flux method, and the measurements were performed on the crystals synthesized in our earlier work \cite{saraswati2026accessing}. The non-magnetic analogue LaGaSi was synthesized to remove the phonon background from NdGaSi in the specific heat data using the gallium self-flux method \cite{gong2024anomalous}. NdAlSi single crystals were grown using the aluminum self-flux method \cite{gaudet2021weyl}. X-ray diffraction (XRD) was carried out on single crystals and powdered single crystals using the SmartLab, Rigaku diffractometer equipped with 9 kW Cu-K$_{\alpha}$ radiation. The obtained XRD data were further refined using the Jana2020 software. To analyze the elemental composition of the obtained crystals, energy-dispersive X-ray spectroscopy (EDS) was performed on a field-emission scanning electron microscope (Quanta 250 FEG).

Single-crystal neutron diffraction (SCND) experiments were conducted on the ISIS Neutron and Muon Source using the SXD diffractometer located at the Rutherford Appleton Laboratory. The SXD instrument operates in the time-of-flight Laue mode, allowing simultaneous coverage of an extensive three-dimensional region of reciprocal space during a single experimental setting \cite{keen2006sxd}. A high-quality single crystal with dimensions of 3  $\times$ 3  $\times$ 3 mm$^3$ was sorted out from the grown batch, and a mass of $\sim$ 150 mg was affixed to an aluminium support pin using thin Al tape strips. The mounted crystal was placed under vacuum and cooled with a closed-cycle cryogenic system. Neutron diffraction datasets were acquired between 5 and 300 K for multiple crystallographic orientations in order to achieve comprehensive reciprocal-space coverage. Subsequent data integration and reduction were performed using the SXD2001 program suite \cite{gutmann2005sxd2001}. For structural refinement, only reflections fulfilling the criterion $I > 3\sigma(I)$ were retained, where $ I$ represents the integrated reflection intensity, and $\sigma (I)$ its corresponding uncertainty. Refinements of both the crystallographic and magnetic structures were subsequently performed using the Jana2020 package. Extinction effects were treated using an isotropic Becker–Coppens type-II formalism \cite{becker1974extinction}, and the corresponding extinction coefficients were refined simultaneously with the structural parameters.

Nd L$_{3}$-edge x-ray absorption spectroscopy (XAS) measurements were carried out on NdGaSi and NdAlSi samples at PETRA III, P64 Advanced XAFS beamline at Deutsches Elektronen-Synchrotron DESY, Hamburg, Germany. The measurements were performed in transmission mode at room temperature on homogeneous pellets (\SI{9}{\mm} diameter and \SI{1}{\mm} thickness). X-ray absorption near-edge structure (XANES) and extended x-ray absorption fine structure (EXAFS) data were collected using a Si(111) monochromator with an energy resolution of $\Delta E/E \sim 10^{-4}$~\cite{caliebe2019high}. The as-obtained data were processed and analysed using the \textsc{Athena} module. The EXAFS spectra were subsequently simulated and analysed using the ARTEMIS software package in the $k$ range of $2$--$10$~\AA$^{-1}$ using the program available within the Demeter software package \cite{newville1995analysis}. This includes background reduction and Fourier transform to derive spectra from the absorption spectra (using ATHENA software), generation of the theoretical EXAFS spectra starting from an assumed crystallographic structure using FEFF6 \cite{newville2001exafs}, and finally fitting of experimental data with the theoretical spectra using ARTEMIS software. To obtain quantitative information about the local environment around the Nd atom in both NdGaSi and NdAlSi, we have analyzed the room-temperature EXAFS spectrum. To account for the oscillatory component in the absorption spectra, the measured absorption coefficient was converted into the EXAFS function, $\chi(E)$, defined as follows \cite{koningsberger1988x}: $\displaystyle \chi(E) = \frac{\mu(E) - \mu_0(E)}{\Delta \mu_0(E_0)}$, where $E_0$ denotes the absorption edge energy, $\mu_0(E)$ represents the bare-atom background contribution, and $\Delta \mu_0(E_0)$ corresponds to the edge-step value at the absorption edge. The energy-dependent absorption coefficient was subsequently transformed into $\chi(k)$ using the relation $\displaystyle k = \sqrt{\frac{2m(E - E_0)}{\hbar^2}}$.

Magnetic measurements were performed using the vibrating sample magnetometer (VSM) equipped with a physical properties measurement system (PPMS, Dynacool, Quantum Design). The specific heat measurements were carried out by the PPMS using the conventional thermal-relaxation method. The CEF analysis was performed using custom code written in Python (Jupyter Notebook), which simultaneously fits the magnetic susceptibility and specific heat data and extracts the CEF parameters, along with the energy eigenvalues and the corresponding eigenvectors. The same code was used to calculate and plot the 4\textit{f} charge density. The code was written with the assistance of Claude Opus 5.5 (Anthropic), and were verified by the authors.

\section{EXPERIMENTAL OBSERVATIONS}
\vspace{3mm}

\begin{figure}[t]
\centering
\includegraphics[width=0.5\textwidth]{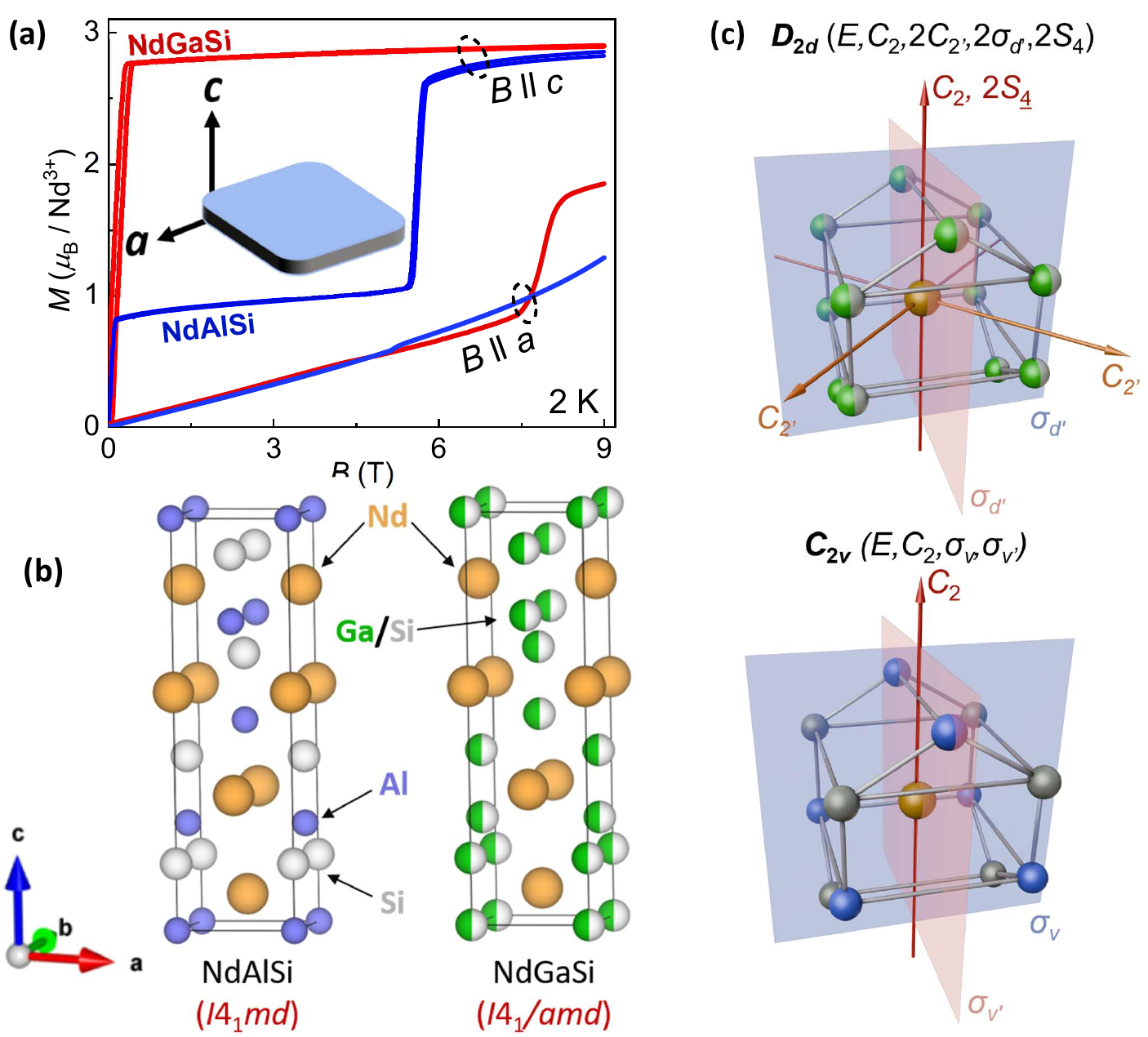}
\caption{(a)  Isothermal magnetisation as a function of magnetic field of both NdGaSi and NdAlSi taken at 2 K, and along both the easy and hard axes.(b) The unit cell of crystalline NdAlSi with space group $I4_1md$ (left), along with that of NdGaSi with $I4_1/amd$ (right) for comparison.  (c) A magnified view of the neighbouring atoms of the Nd-ion for both NdGaSi (top) and NdAlSi (bottom), along with their respective point group symmetry elements.}
\label{fig1}
\end{figure}

\subsection{Magnetic and structural properties}

The temperature-dependent susceptibility $\chi(T)$ shows an anomaly at $T_\mathrm{C}=11$~K for fields both in and out of plane, marking the onset of magnetic order (see the supplementary information (SI) \cite{supply}). Figure~\ref{fig1}a compares the isothermal magnetization $M(B)$ of NdGaSi and NdAlSi at 2~K. For NdGaSi with $B\parallel c$ (easy axis), $M(B)$ rises sharply, shows an inflection near $0.3$~T, and saturates at $2.9\,\mu_\mathrm{B}$, close to the free-ion value $g_JJ\mu_\mathrm{B}\approx3.27\,\mu_\mathrm{B}$, as expected for a collinear FM with strong uniaxial anisotropy. NdAlSi instead shows an MMT near $0.1$~T along its easy axis; neutron diffraction associates this with the amplitude-modulated INC order, $\mathbf{k}=(\frac{2}{3}+\delta,\frac{2}{3}+\delta,0)$, below 7.2~K, which locks into a commensurate canted ``up-down-down'' FIM along $c$ below 3.3~K \cite{gaudet2021weyl}. Above 0.1~T, $M(B)$ rises gradually and shows a second MMT at $\sim5$~T, where the moments become fully polarized.

Along the hard axis, the contrast is equally clear. For NdGaSi with $B\parallel a$, $M(B)$ increases almost linearly up to $7.9$~T, where a sharp first-order magnetic transition (FOMT) occurs, and continues to rise without saturating up to 9~T \cite{asti1990first,lyakhova2017analysis,gorbunov2016first}. The critical field $B_\mathrm{c}=7.9$~T directly measures the large anisotropy barrier that must be overcome to rotate the moments away from $c$ \cite{asti1990first}. For NdAlSi, $M(B)$ along the hard direction is featureless and does not saturate up to 9~T. These contrasting responses reflect different GS anisotropies and exchange frustration, which we trace below to the distinct Nd-site symmetries and CEF schemes.

NdGaSi crystallises in the centrosymmetric space group $I4_1/amd$ ($\alpha$-ThSi$_2$ type, No.~141) with complete Ga/Si antisite disorder on the single $8e$ site [Fig.~\ref{fig1}b, right] \cite{darone2013rare,saraswati2026accessing}. NdAlSi instead adopts the noncentrosymmetric $I4_1md$ structure (LaPtSi type, No.~109), an ordered derivative of $\alpha$-ThSi$_2$ in which Al and Si occupy two distinct $4a$ sites [Fig.~\ref{fig1}b, left]. This ordering removes the inversion centre and makes the structure polar along $c$. Because Al and Si have very similar x-ray scattering factors \cite{hubbell1975atomic}, the NdAlSi structure could be established only by combining SCND and second-harmonic generation \cite{gaudet2021weyl}, whereas the clear contrast between Ga and Si makes the NdGaSi structure readily resolvable by single-crystal XRD.

The Nd coordination also reflects this symmetry difference, as illustrated in Fig.~\ref{fig1}c. In NdAlSi (bottom panel), Nd occupies the $4a$ Wyckoff position with a local point symmetry of $C_{2v}$ ($2mm$). This symmetry includes a two-fold rotation axis along the $c$ direction and two vertical mirror planes ($m$ and $m'$) that intersect along that axis, but lacks an inversion centre. In contrast, NdGaSi (top panel) places the Nd$^{3+}$ ion at the $4b$ Wyckoff position, where the local point group is the higher tetragonal symmetry $D_{2d}$ ($\bar{4}m2$). This group features an improper four-fold rotation axis ($\bar{4}$) along $c$,  two independent two-fold axes perpendicular to $c$, and two diagonal mirror planes ($m$).  The presence of the $S_4$ improper rotation forbids all $m=\pm2$ Stevens operators (discussed later), making the $D_{2d}$ site symmetry the key distinction from the $C_{2v}$ symmetry in NdAlSi. 
\begin{table*}[htbp]
\centering
\caption{Symmetry-allowed magnetic space groups and corresponding irreducible representations obtained from representational analysis of the paramagnetic $I4_1/amd$ structure with propagation vector $\mathbf{k}=(0\ 0\ 0)$, together with the symmetry-permitted magnetic moment components at the Nd sites.}
\vspace{2mm}
\label{tab:magnetic_space_groups}
\begin{ruledtabular}
\begin{tabular}{*{12}{c}}

Shubnikov space group & Irreducible representation (IR) & Moment & Global moment \\
\midrule
$I4_1/amd$               & $\Gamma_1^+$ & $0\;0\;0$ & $0\;0\;0$ \\
$I4_1'/amd'$             & $\Gamma_2^+$ & $0\;0\;0$ & $0\;0\;0$ \\
$I4_1/am'd'$             & $\Gamma_3^+$ & $0\;0\;M$ & $0\;0\;M$ \\
$I4_1'/am'd$             & $\Gamma_4^+$ & $0\;0\;0$ & $0\;0\;0$ \\
$I4_1/a'm'd'$            & $\Gamma_1^-$ & $0\;0\;0$ & $0\;0\;0$ \\
$I4_1'/a'm'd$            & $\Gamma_2^-$ & $0\;0\;M$ & $0\;0\;0$ \\
$I4_1/a'md$              & $\Gamma_3^-$ & $0\;0\;0$ & $0\;0\;0$ \\
$I4_1'/a'md'$            & $\Gamma_4^-$ & $0\;0\;0$ & $0\;0\;0$ \\
$Imm'a'$                 & $\Gamma_5^+$ & $0\;M\;0$ & $0\;M\;0$ \\
$Im'ma$                  & $\Gamma_5^-$ & $M\;0\;0$ & $0\;0\;0$ \\
$Fdd'd'$                 & $\Gamma_5^+$ & $M\;M\;0$ & $M\;M\;0$ \\
$Fd'dd$                  & $\Gamma_5^-$ & $M\;M\;0$ & $0\;0\;0$ \\
$C2'/c'$                 & $\Gamma_5^+$ & $M\;M\;0$ & $M\;M\;0$ \\
$C2'/c$                  & $\Gamma_5^-$ & $M\;M\;0$ & $0\;0\;0$ \\

\end{tabular}
\end{ruledtabular}
\end{table*}

\begin{table}[t]
\centering
\caption{Results obtained from magnetic refinement of SCND data for NdGaSi at 5 K following $I4_1/am'd'$ Shubnikov space group. Here, $m_a$, $m_b$, and $m_c$ represent the component of magnetic moment of Nd along crystallographic $a$, $b$ and $c$ axes respectively, while total magnetic moment is denoted by $M$.}
\vspace{2mm}
\label{tab:refinement_results}
\begin{ruledtabular}
\begin{tabular}{l c}

Refinement parameter & Value \\
\midrule
$m_a$ ($\mu_\mathrm{B}$) & 0 \\
$m_b$ ($\mu_\mathrm{B}$) & 0 \\
$m_c$ ($\mu_\mathrm{B}$) & 3.031 \\
$M$ ($\mu_\mathrm{B}$) & 3.031 \\
\midrule
GoF & 6.27 \\
$R_{\text{all}}$ & 8.93 \\
$wR_{\text{all}}$ & 22.51 \\
$R_{\text{mag}}$ & 13.42 \\

\end{tabular}
\end{ruledtabular}
\end{table}

\begin{figure}[t]
\centering
\includegraphics[width=0.5\textwidth]{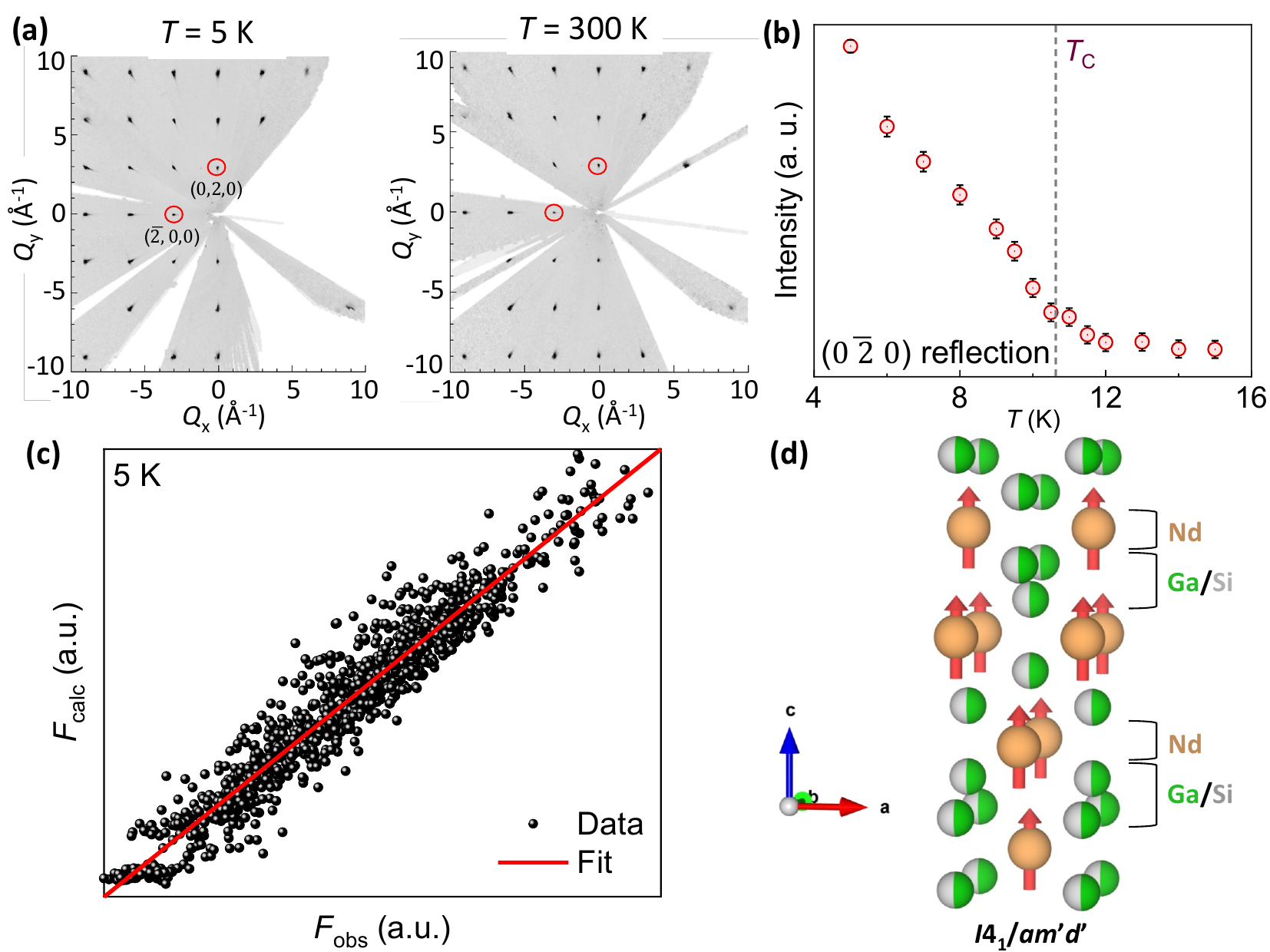}
\caption{(a) SCND intensity maps of NdGaSi in the (\textit{h k} 0) layer recorded at 5 K and 300 K. (b) Integrated intensity vs T plot for ($0~\bar{2}~0$) magnetic reflection clearly indicating the onset of magnetic ordering. (c) Calculated vs observed structure factors ($F_{cal}$ vs $F_{obs}$) at 5 K for $I4_1/am'd'$ model.  (d) Ground-state magnetic structure at 5 K as derived from the SCND data.}
\label{fig3}
\end{figure}

\subsection{Single crystal neutron diffraction}
SCND measurements were performed on NdGaSi over the temperature range 5--300 K to determine its magnetic GS configuration and the orientation of the ordered Nd moments. Figure~\ref{fig3}a displays the reciprocal-space neutron-diffraction intensity maps recorded in the $(h~k~\mathrm{0})$ scattering plane at 300 and 5 K. At 300 K, all the observed reflections can be indexed as nuclear Bragg peaks allowed by the centrosymmetric tetragonal space group \(I4_{1}/amd\), without any detectable contribution from magnetic scattering. Upon cooling below the magnetic ordering temperature, \(T_{\mathrm{C}}\sim 11\) K, pronounced intensity enhancements develop at several pre-existing nuclear Bragg positions, as marked by the red circles in Fig.~\ref{fig3}a. These enhancements arise from the additional magnetic contribution to the neutron-scattering cross section and provide direct microscopic evidence for the development of long-range magnetic order. Importantly, no additional magnetic superlattice reflections were detected within the measured reciprocal-space volume. Instead, the magnetic scattering remains superimposed on the nuclear Bragg reflections. This observation establishes that the magnetic order is commensurate with the underlying crystallographic lattice and can be described by the propagation vector \(\mathbf{k}=(0,0,0)\). The magnetic and crystallographic unit cells, therefore, coincide, excluding an INC modulation or an AFM enlargement of the unit cell within the sensitivity of the present experiment. The absence of satellite reflections is particularly significant in the present context because the structurally related compound NdAlSi develops an INC magnetic order before transforming into a commensurate FIM state at lower temperatures. In contrast, NdGaSi develops a commensurate collinear FM structure below \(T_{\mathrm{C}}\), demonstrating that the Ga/Si-disordered centrosymmetric structure stabilizes a magnetic GS fundamentally different from the INC helical order observed in NdAlSi \cite{gaudet2021weyl}.

To further test the crystal structure, the 300~K neutron data were refined with both the disordered ($I4_1/amd$) and ordered ($I4_1md$) models. The disordered model gives better agreement with the experimental data (see SI \cite{supply}), while the ordered model yields a non-positive thermal ellipsoid for the Si site, an unphysical result. This suggests that NdGaSi crystallizes into the Ga/Si disordered structure bearing the space group $I4_1/amd$.

To investigate the thermal evolution of the magnetically ordered state, the integrated intensities of selected Bragg reflections containing appreciable magnetic contributions were followed as a function of temperature. In particular, the integrated intensity of the \((0\,\bar{2}\,0)\) reflection [Fig.~\ref{fig3}b] exhibits a systematic enhancement below \(T_{\mathrm{C}}\), directly reflecting the development of the ordered Nd moment. For magnetic neutron scattering, only the component of the magnetic moment perpendicular to the scattering vector contributes to the intensity. Since the scattering vector of the \((0\,\bar{2}\,0)\) reflection lies along the crystallographic \(b\) direction, its pronounced magnetic enhancement is consistent with an ordered-moment component perpendicular to \(b\), as subsequently determined to lie entirely along the \(c\)-axis from the magnetic-structure refinement. The onset temperature of the intensity enhancement agrees well with the anomaly observed in the bulk magnetic susceptibility \(\chi(T)\) [Fig.~S2a of the SI \cite{supply}], confirming the magnetic origin of the additional low-temperature scattering. It should be noted that the magnetic contribution to the Bragg intensity is proportional to the square of the ordered moment. Consequently, the temperature dependence of the \((0\,\bar{2}\,0)\) intensity provides a microscopic measure of the evolution of the magnetic order parameter below \(T_{\mathrm{C}}\). The simultaneous appearance of the magnetic Bragg intensity and the bulk susceptibility anomaly demonstrates that the transition near 11 K corresponds to the establishment of a spatially coherent long-range magnetic state rather than merely the development of short-range FM correlations.

To determine the symmetry and orientation of the ordered Nd moments, magnetic representational analysis was performed using the \textsc{Jana2020} software package. For the \(\mathbf{k}=(0,0,0)\) propagation vector, the little group of \(\mathbf{k}\) retains the symmetry operations of the parent crystallographic space group. The magnetic representation associated with the Nd \(4b\) Wyckoff site was decomposed into the irreducible representations of this little group. Each irreducible representation defines a set of symmetry-adapted basis functions that specifies the relative orientations and phase relationships of the magnetic moments on the symmetry-equivalent Nd sites. Depending on the active irreducible representation and the corresponding direction of the magnetic order parameter in representation space, the symmetry-adapted configurations generate distinct magnetic space groups. These magnetic space groups impose specific restrictions on the allowed moment components and distinguish configurations with a net FM moment from those in which the moments compensate within the crystallographic unit cell. The representational analysis therefore provides a systematic symmetry classification of all magnetic structures compatible with both the parent \(I4_{1}/amd\) structure and the experimentally established propagation vector \(\mathbf{k}=(0,0,0)\). The symmetry-allowed magnetic models, together with their corresponding irreducible representations, magnetic space groups, local moment components, and net magnetic moments, are summarized in Table~\ref{tab:magnetic_space_groups}.

Among the 14 symmetry-allowed magnetic space groups, the \(I4_{1}/am'd'\) model is the only tetragonal configuration that permits a uniaxial FM moment strictly along the crystallographic \(c\) axis. This model preserves the tetragonal character of the parent structure while allowing the Nd moments to transform as an FM order parameter parallel to \(c\). By contrast, the remaining symmetry-allowed models either forbid magnetic ordering on the Nd site, produce compensated magnetic structures with zero net moment, permit moments within the basal plane, or lower the magnetic symmetry from tetragonal to orthorhombic or monoclinic. 
Rietveld refinements of the 5 K SCND data were performed using the candidate magnetic structures obtained from the representational analysis. Among these models, the \(I4_{1}/am'd'\) magnetic space group provides the best agreement between the observed and calculated magnetic intensities. The quality of the refinement is further demonstrated by the comparison between the observed and calculated structure factors at 5 K in Fig.~\ref{fig3}c. The refinement yields vanishing in-plane moment components, \(m_a=m_b=0\), together with an ordered moment of \(m_c=3.031\,\mu_{\mathrm{B}}\) per Nd ion along the \(c\) axis. Thus, no measurable canting of the Nd moments away from the tetragonal axis is required to reproduce the experimental diffraction intensities.

The refined ordered moment is close to the saturation magnetization of approximately \(2.9\,\mu_{\mathrm{B}}\) per Nd ion obtained from the \(M(B)\) measurement for \(B\parallel c\). The small difference between the two values can reasonably arise from the distinct experimental conditions and from uncertainties associated with the magnetic-intensity and bulk-magnetization analyses. More importantly, their close agreement demonstrates the consistency between the microscopic ordered moment determined by SCND and the macroscopic FM moment measured by magnetization. The absence of any refined in-plane component is also consistent with the nearly linear and nonsaturating \(M(B)\) response for a field applied perpendicular to the \(c\) axis, further supporting a strong uniaxial magnetic anisotropy. The resulting GS magnetic structure at 5 K is illustrated in Fig.~\ref{fig3}d, and the corresponding refinement parameters are listed in Table~\ref{tab:refinement_results}. It consists of a collinear FM arrangement in which all Nd moments are aligned parallel to the crystallographic \(c\) axis, without any detectable in-plane canting or spatial modulation. This microscopic magnetic configuration is fully consistent with the strong easy-axis anisotropy observed in \(\chi(T)\) and \(M(B)\) [Figs.~S2(a) and \ref{fig1}a, respectively]. Therefore, the combined reciprocal-space mapping, temperature-dependent Bragg-intensity analysis, magnetic representational analysis, and refinement of the ordered-state SCND data establish that NdGaSi hosts a commensurate, collinear, and strongly uniaxial FM GS characterized by \(\mathbf{k}=(0,0,0)\) and Nd moments aligned along the tetragonal \(c\)-axis.

\begin{figure}[t]
\centering
\includegraphics[width=0.5\textwidth]{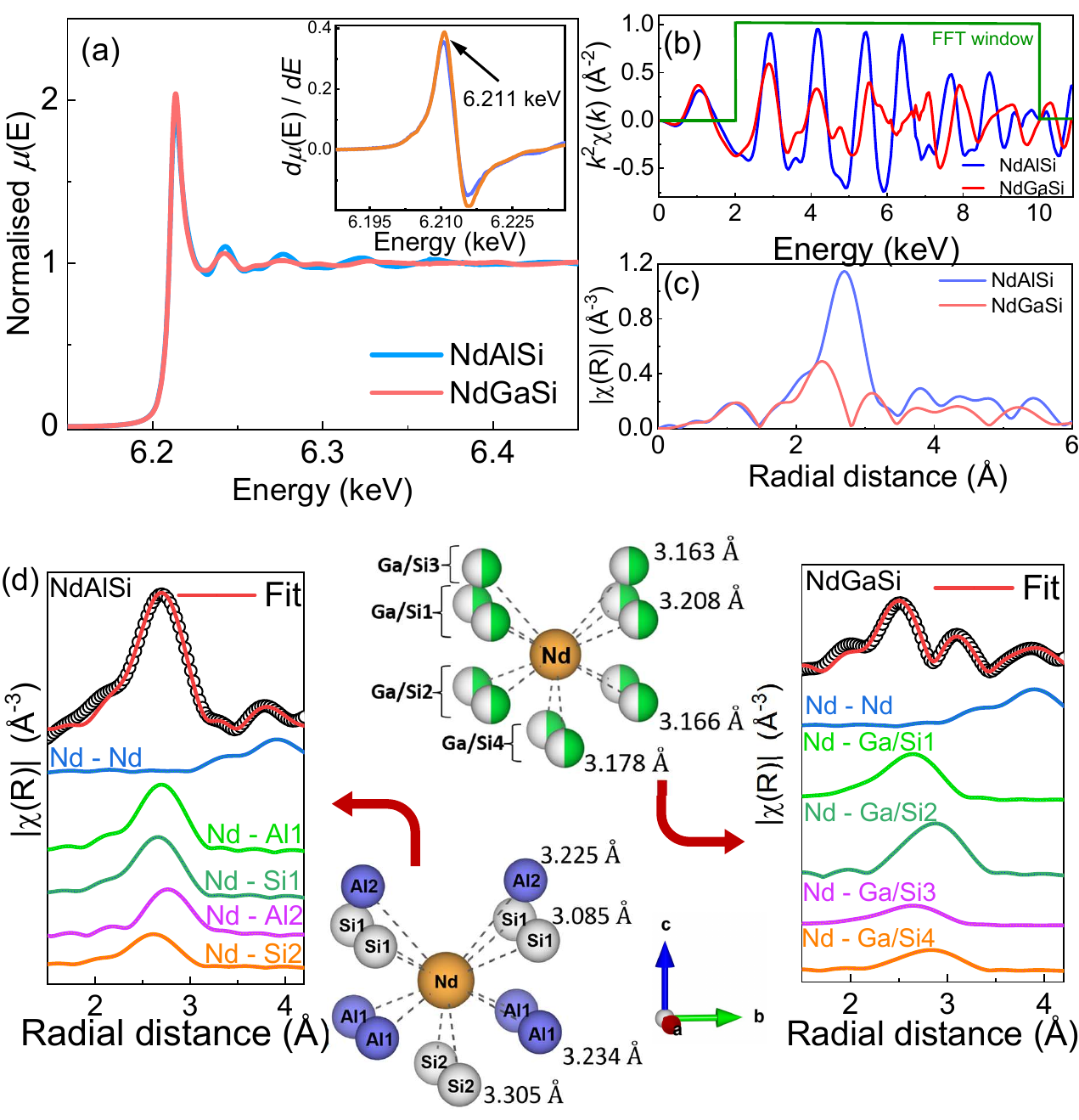}
\caption{(a) Normalised Nd-L$_3$ XANES spectra of NdGaSi and NdAlSi. The inset represents the first derivative of the spectra, highlighting the absorption edge energy ($E_0$). (b) The EXAFS spectra plotted for both systems in \textit{k}-space, and (c) their corresponding Fourier transformations to \textit{R}-space. (d) Fitting performed on the EXAFS spectra in \textit{R}-space for NdAlSi, along with the schematic representation of the local environment obtained from the fit (Left panel). Corresponding fitting and the local environment obtained from the fit are also displayed for NdGaSi (Right panel). Individual scattering paths are vertically shifted for visual clarity.}
\label{exafs}
\end{figure}

\vspace{1mm}

\subsection{X-ray absorption spectroscopy}

Nd $L_3$-edge XAS, a direct probe of the local environment, was used to help distinguish the Nd point groups of NdAlSi and NdGaSi. Figure~\ref{exafs}a shows the normalized XANES spectra for both compounds, along with their corresponding first derivatives in the inset. The absorption edge positions, determined from the maxima of the first derivatives, are nearly identical for both systems at $\sim$  6.211 keV (Nd-L$_3$ edge), confirming the Nd$^{3+}$ valence state and ruling out any significant oxidation or valence fluctuation between the two compounds \cite{choi2007nd,kanthal2025dysprosium}. The overall shape of the XANES spectra, however, exhibits noticeable differences, reflecting the distinct local coordination environments of Nd in both systems.

\begin{table*}[!htbp]
\caption{EXAFS fitting parameters obtained for NdAlSi and NdGaSi. }
\label{tab:exafs_comparison}
\renewcommand{\arraystretch}{1.2}
\setlength{\tabcolsep}{2.5pt}
\begin{ruledtabular}
\begin{tabular}{l c c c c c c c}

\multicolumn{4}{c}{\textbf{NdAlSi}} & \multicolumn{4}{c}{\textbf{NdGaSi}} \\
\hline
\textbf{Path} & \textbf{$R$ (\AA)} & \textbf{$N$} & \textbf{$\sigma^2$ (\AA$^2$)} &
\textbf{Path} & \textbf{$R$ (\AA)} & \textbf{$N$} & \textbf{$\sigma^2$ (\AA$^2$)} \\
\hline
Nd--Nd  & $4.215 \pm 0.016$ & 8 & $0.014 \pm 0.0017$ &
Nd--Nd  & $4.182 \pm 0.009$ & 8 & $0.02194 \pm 0.001$ \\
Nd--Al1 & $3.234 \pm 0.004$ & 4 & $0.00323 \pm 0.0007$ &
Ga/Si1  & $3.208 \pm 0.005$ & 4 & $0.01098 \pm 0.0006$ \\
Nd--Si1 & $3.085 \pm 0.006$ & 4 & $0.00759 \pm 0.0009$ &
Ga/Si2  & $3.163 \pm 0.007$ & 4 & $0.01565 \pm 0.001$ \\
Nd--Al2 & $3.225 \pm 0.008$ & 2 & $0.00136 \pm 0.001$ &
Ga/Si3  & $3.166 \pm 0.013$ & 2 & $0.01395 \pm 0.002$ \\
Nd--Si2 & $3.305 \pm 0.011$ & 2 & $0.00473 \pm 0.0016$ &
Ga/Si4  & $3.178 \pm 0.017$ & 2 & $0.01821 \pm 0.003$ \\

\end{tabular}
\end{ruledtabular}
\end{table*}

The EXAFS oscillations, extracted after standard background subtraction and normalization, are presented in $k$-space in Fig.~\ref{exafs}b. NdAlSi exhibits significantly stronger EXAFS oscillations compared to NdGaSi. This key difference is a direct consequence of the local structural order. In ordered NdAlSi, the well-defined Al and Si sublattices produce coherent scattering from distinct atomic shells, preserving the EXAFS amplitude. In contrast, the random Ga/Si site mixing in NdGaSi introduces static disorder, which attenuates the photoelectron wave and reduces the oscillation amplitude \cite{yamazaki2000defect,keiber2013complex,malik2022antisite,malik2022role}. The Fourier transforms (FT) of the \textit{k}-space data were performed over a range of 2--10 Å$^{-1}$ to obtain the $R$-space as shown in Fig.~\ref{exafs}c. The FT magnitude for NdAlSi shows a prominent, well-defined first-shell peak, whereas the peak for NdGaSi is significantly less intense and comparatively broader. This reduction in the FT peak height is a crucial hint regarding the static disorder arising from the Ga/Si antisite mixing.

The key structural difference between the two compounds lies in the local environment of the Nd$^{3+}$ ion. In NdAlSi, the Al and Si atoms occupy distinct crystallographic sites, resulting in dissimilar Nd--Al and Nd--Si bond lengths. This reflects the broken inversion symmetry and the noncentrosymmetric local environment ($C_{2v}$). In contrast, in NdGaSi, the Ga and Si atoms are randomly mixed on a single crystallographic site, giving highly similar bond lengths. This is the signature of the averaged, higher-symmetry ($D_{2d}$) Nd environment of the centrosymmetric structure. To quantitatively verify this structural difference at the local level, the EXAFS data were fitted using the ARTEMIS software package \cite{ravel2005athena}. The theoretical scattering paths were generated using the FEFF6 code, based on the crystallographic structures obtained from single-crystal XRD refinements \cite{saraswati2026accessing,gaudet2021weyl}. For NdAlSi, the ordered noncentrosymmetric $I4_1md$ structure was used, with Al and Si occupying distinct $4a$ Wyckoff sites. For NdGaSi, the centrosymmetric $I4_1/amd$ structure with 50/50 Ga/Si mixing on the 8e Wyckoff site was employed as the starting model. Figure~\ref{exafs}d shows the best-fit $\chi(R)$ versus $R$ spectra along with the experimental data for both compounds, with individual scattering paths vertically shifted for visual clarity. The fitting was performed keeping the coordination number (\textit{N}) fixed to the crystallographic value. The refined structural parameters are summarized in Table~\ref{tab:exafs_comparison}. For NdAlSi, the fit yields four distinct Nd-to-ligand bond lengths. The total spread between the shortest and longest bonds is approximately 0.22 \AA. This substantial spread directly reflects the ordered nature of the Al and Si sublattices, where the two elements occupy distinct crystallographic sites with different average distances from the central Nd$^{3+}$ ion. The corresponding Debye-Waller factors are consistently low ($\sigma^2 = 0.0014$--$0.0076$ \AA$^2$), indicating that the individual shells are well-defined and exhibit minimal static disorder, and confirming the noncentrosymmetric local environment at the Nd site.

\vspace{2mm}

In striking contrast, the EXAFS fit for NdGaSi reveals a qualitatively different local structure. The refined bond lengths are highly similar, with a total spread of only 0.045 \AA. This collapsed distribution of distances is a direct consequence of the random Ga/Si site mixing, where the two elements share a single crystallographic site. The Debye-Waller factors for NdGaSi are substantially larger ($\sigma^2 = 0.010$--$0.018$ \AA$^2$) compared to NdAlSi. This significant increase reflects the static disorder introduced by the random distribution of Ga and Si atoms, which creates local bond-length fluctuations around the average value \cite{piskorska2026random,fang1983exafs,martinez2025mixed,shuvaeva2024mossbauer}.

\vspace{2mm}

The EXAFS results, thus, confirm the structural picture established above: The ordered Al/Si sublattice in NdAlSi produces a wide spread of Nd--ligand bond lengths (0.22 \AA), reflecting the broken inversion symmetry and $C_{2v}$ local symmetry, while the random Ga/Si mixing in NdGaSi collapses the bond lengths to a narrow distribution (0.045 \AA), directly reflecting the restoration of the inversion center and the $D_{2d}$ environment. The substantially larger Debye-Waller factors in NdGaSi provide quantitative evidence for the static disorder arising from the antisite mixing, complementing our SCND analysis. These structural differences provide the necessary input for the CEF analysis that follows, where we examine how the distinct point group symmetries of Nd in the respective lattices manifest in the CEF level schemes and, ultimately, the magnetic GS of the two compounds.

\begin{figure*}
\centering
\includegraphics[width=1.0\textwidth]{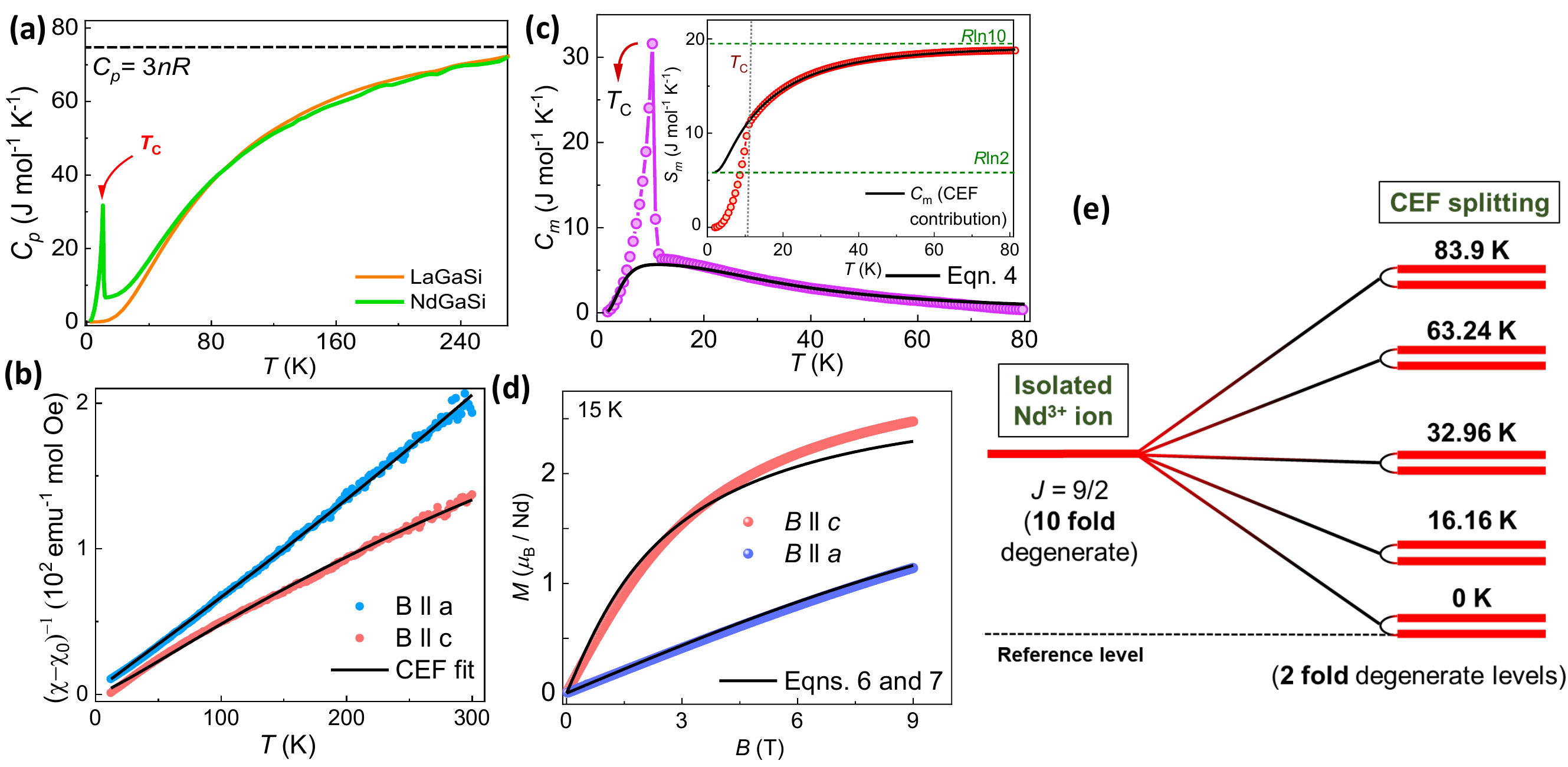}
\caption{(a) Specific heat as a function of temperature plotted for both NdGaSi and LaGaSi, indicating the transition temperature. (b) $(\chi -\chi_0)^{-1}$ plotted as a function of temperature for the field-cooled mode with \textit{B } = 0.02 T. Black lines indicate the CEF fit (Eqn. 3). (c) Magnetic specific heat as a function of temperature. The inset shows the magnetic entropy as a function of temperature. Black lines indicate the respective physical quantities obtained from the CEF fit. (d) $M(B)$ data of NdGaSi at 15 K. Black lines indicate the one obtained from the CEF fit. (e) Representative figure of the obtained CEF scheme of NdGaSi.} 
\label{fig4}
\vspace{-0.5cm}
\end{figure*}

\vspace{3mm}
\section{CEF ANALYSIS AND DISCUSSION}

\vspace{3mm}
\subsection{CEF analysis }

The SCND result in the previous section establishes that NdGaSi orders ferromagnetically below $T_{\mathrm{C}} \sim 11$ K with a collinear spin structure along the $c$-axis [$\mathbf{k} = (0,0,0)$], indicating a strong uniaxial anisotropy. To understand the microscopic origin of this anisotropy, we now examine the CEF acting on the Nd$^{3+}$ ions.

The CEF acting on a rare-earth system lifts the degeneracy of the Hund's-rule multiplet and, most importantly, fixes the GS wavefunction that sets the single-ion anisotropy \cite{burns1962shielding,trammell1963magnetic,crow2012crystalline,guertin2012crystalline}. Stevens introduced the operator-equivalent formalism \cite{stevens1952matrix}, in which the CEF Hamiltonian is written as $\mathcal{H}_{\rm CEF}=\sum_{l,m}B_l^mO_l^m$, with Stevens operators $O_l^m$ and CEF parameters $B_l^m$, and Hutchings related these parameters to the charges of the surrounding ligands within the point-charge model (PCM) \cite{hutchings1964point}. In practice, the symmetry-allowed $B_l^m$ can be determined by fitting experimental data. This approach describes the susceptibility, specific heat, magnetization and inelastic neutron scattering (INS) of many rare-earth intermetallics \cite{bauer2009magnetism,fulde1985magnetic,uzoh2023influence,jin2024easy,takeuchi2003anisotropic,banda2018crystalline}. For \textit{R}AlSi, Yang \textit{et al.} determined the GS wavefunctions and level schemes of polycrystals from INS, heat capacity and susceptibility \cite{yang2025crystalline}, while only qualitative analyses exist for \textit{R}Ga\textit{X} \cite{ram2024crystalline,ram2023magnetic,swami2025crystalline,swami2025crystallinepgs}.

For Nd$^{3+}$ ($4f^3$), Hund's rules give a $^4I_{9/2}$ ground multiplet with $J=9/2$, well separated from the first excited multiplet ($^4I_{11/2}$) by spin--orbit coupling. Within this multiplet, the CEF lifts the $(2J+1)=10$-fold degeneracy and, because Nd$^{3+}$ is a Kramers ion, splits it into five Kramers doublets \cite{waltercef,gopal2012specific}. The composition of the lowest doublet in terms of the $|J,M\rangle$ states determines the size and direction of the single-ion moment, and hence the single-ion anisotropy, while inter-ionic exchange acts on this doublet to produce long-range order below $T_\mathrm{C}$.

In the PCM \cite{hutchings1964point}, each surrounding ion is replaced by a point charge $q_i$ at its crystallographic position $r_i$, and the electrostatic potential $V(r)$ at the Nd$^{3+}$ site is expanded in spherical harmonics. Averaging over the $4f$ charge cloud with Stevens' operator equivalents \cite{stevens1952matrix} gives the CEF Hamiltonian of the $J$ multiplet as

\begin{equation}
\mathcal{H}_{\mathrm{CEF}} = \sum_{l=2,4,6} \sum_{m=-l}^{l} B_l^m O_l^m,
\end{equation}
where $O_l^m$ are Stevens operators, and $B_l^m$ are CEF parameters. Here, the symmetry-allowed $B_l^m$ are treated as free parameters and determined from experimental observables.

\subsubsection{CEF Hamiltonian of NdGaSi}

The single-crystal XRD analysis in our earlier work \cite{saraswati2026accessing}, combined with neutron diffraction and XAS analysis earlier in the paper, confirms that NdGaSi crystallises in the centrosymmetric space group $I4_1/amd$ with complete Ga/Si disorder. The Nd$^{3+}$ ions occupy the $4b$ Wyckoff position, which has tetragonal point symmetry $D_{2d}$ ($\bar{4}m2$). Applying the restrictions under $D_{2d}$, the CEF Hamiltonian reduces to \cite{yang2025crystalline,bauer2009magnetism,waltercef}
$$
\mathcal{H}_{\mathrm{CEF}}^{D_{2d}} = B_2^0 O_2^0 + B_4^0 O_4^0 + B_4^4 O_4^4 + B_6^0 O_6^0 + B_6^4 O_6^4.
$$
The sixth-order terms (\textit{l} = 6) were fixed to zero to reduce the number of free parameters. Diagonalising $\mathcal{H}_{\mathrm{CEF}}$ yields $10$ eigenvalues $E_n$ and their corresponding eigenstates $|\Psi_n\rangle$ in the form of five Kramers doublets (time-reversal pairs). 

The single-ion susceptibility (\textit{i = x, y, z} …), comprising Curie and Van Vleck terms, is given by
\begin{multline}
\chi_{i}^\mathrm{CEF}(T) = \frac{N_A (g_J\mu_\mathrm{B})^2}{Z} \biggl[ 
\sum_n \frac{|\langle \Psi_n | J_i | \Psi_n \rangle|^2}{k_\mathrm{B} T} e^{-\frac{E_{n}}{k_\mathrm{B} T}} \qquad  \\
+ \sum_{E_n \neq E_m} |\langle \Psi_m | J_i | \Psi_n \rangle|^2 \frac{e^{-\frac{E_{n}}{k_\mathrm{B} T}} - e^{-\frac{E_{m}}{k_\mathrm{B} T}}}{E_m - E_n} 
\biggr].
\end{multline}
where $Z = \sum_n e^{-E_n/k_B T}$ is the partition function \cite{bauer2009magnetism,ram2023magnetic}. The first (Curie) term comes from matrix elements within each degenerate doublet, i.e.\ the moment the doublet already carries, whereas the second (Van Vleck) term describes the moment induced by field-driven mixing with excited CEF levels. The molecular field ($\lambda$) is incorporated into the final equation, along with subtracting any temperature-independent component ($\chi_0$), to obtain the final form:

\begin{equation}
[\chi_{i}(T) - \chi_{0}]^{-1} = \bigl[ \chi_{i}^{\mathrm{CEF}}(T) \bigr]^{-1} - \lambda_{i}
\end{equation}

The hump in specific heat ($C_\mathrm{P}$) arises from the thermal population of the CEF levels known as the Schottky anomaly \cite{gopal2012specific,waltercef} and is represented as:
\begin{multline}
  C_{CEF}(T) = 
   \left(\frac{R}{T^2}\right)\\\left[\frac{\sum_{n}^{ } g_{n}e^{-\frac{E_{n}}{k_\mathrm{B} T}} \sum_{n}^{ } g_{n}{E_{n}}^2e^{-\frac{E_{n}}{k_\mathrm{B} T}}-\left(\sum_{n}^{ } g_{n}E_ne^{-\frac{E_{n}}{k_\mathrm{B} T}} \right)^2}{\left(\sum_{n}^{ } g_{n}e^{-\frac{E_{n}}{k_\mathrm{B} T}} \right)^2}\right]
\end{multline}
with degeneracy $g_n=2$ for each doublet. The phonon background was subtracted using the specific heat of isostructural, non-magnetic LaGaSi.

\begin{table}[t]
\centering
\caption{CEF parameters obtained for NdGaSi.}
\label{tab:CEFparams}
\renewcommand{\arraystretch}{1.2}
\begin{ruledtabular}
\begin{tabular}{l c}

Parameter & Value \\
\hline
$B_2^0$ (K)  & $-0.775$ \\
$B_4^0$ (K)  & $0.015$ \\
$B_4^4$ (K)   & $0.058$ \\
\hline
$\lambda_c$ (mol/emu)   & $1.539$ \\
$\lambda_{ab}$ (mol/emu) & $1.113$ \\

\end{tabular}
\end{ruledtabular}
\end{table}

\begin{table*}[!htbp]
\caption{Calculated CEF energy levels (in K) and eigenvectors for NdGaSi in the $|J=9/2, M_J\rangle$ basis.}
\label{tab:eigen}
\renewcommand{\arraystretch}{1.2}
\setlength{\tabcolsep}{2.5pt}
\begin{ruledtabular}
\begin{tabular}{*{12}{c}}

 $E$ (K) & $|-\frac{9}{2}\rangle$ & $|-\frac{7}{2}\rangle$ & $|-\frac{5}{2}\rangle$ & $|-\frac{3}{2}\rangle$ & $|-\frac{1}{2}\rangle$ & $|\frac{1}{2}\rangle$ & $|\frac{3}{2}\rangle$ & $|\frac{5}{2}\rangle$ & $|\frac{7}{2}\rangle$ & $|\frac{9}{2}\rangle$ \\
\hline
 0.00    & 0& -0.987&0 &0 & 0&0.163 &0 &0 & 0 & -0.037\\
 0.00   & 0.037& 0& 0&0 &-0.163 &0 &0 &0 &0.987 &0 \\
16.16  &0&0 &0 &0.364 &0 &0 & 0&-0.931 &0 &0 \\
16.16  &0 &0 &0.931 &0 &0 &0 &-0.364 &0 &0 &0 \\
32.96  &0.987 &0 &0 &0 &-0.145 &0 &0 &0 & -0.062&0 \\
32.96  &0 &-0.062 & 0&0 & 0&-0.145 &0 &0&0 &0.987\\
63.24  &0 &0 &-0.365 &0 &0 &0 &-0.931 &0 &0 &0 \\
63.24  &0 &0 &0 &0.931 &0 &0 &0 &0.365 &0 &0 \\
83.90  &-0.154 &0 &0 &0 &-0.976 &0 &0 &0 &-0.156 &0 \\
83.90  &0 &0.156 &0 &0 &0 &0.976 &0 &0 &0 &0.154 \\

\end{tabular}
\end{ruledtabular}
\end{table*}
\vspace{3mm}

\subsubsection{Simultaneous fitting and results}

Figure~\ref{fig4}a shows the specific heat $C_p(T)$ of NdGaSi and LaGaSi. The sharp $\lambda$-type anomaly at $T_\mathrm{C}=11$~K marks the bulk FM transition seen in $\chi(T)$ and SCND (Sec.~IIIB). Subtracting $C_p(T)$ of LaGaSi gives the magnetic specific heat $C_\mathrm{m}(T)$ [Fig.~\ref{fig4}c], which shows a broad Schottky anomaly from the thermal population of excited CEF doublets. The magnetic entropy $S_\mathrm{m}(T)$ (inset), obtained by integrating $C_\mathrm{m}/T$, reaches $R\ln2$ below $T_\mathrm{C}$ and saturates at $R\ln10$ near 100~K, so all five doublets lie within $\sim90$~K and the low-temperature magnetism is governed by the $R\ln2$ GS doublet.

\vspace{2mm}

To avoid overfitting and unphysical parameters, we have simultaneously fitted the $\chi(T)$ along both \textit{c}-and \textit{a}-directions (Fig \ref{fig4}b) using Eq. 3, together with $C_{\mathrm{m}}(T)$ (Fig. \ref{fig4}c) using Eq. 4 above $T_\mathrm{C}$, to avoid contributions from exchange interactions, using a custom Python code with a global fitting algorithm. A CEF scheme obtained this way must reproduce both observables, which excludes parameter sets that fit one but fail the other, as detailed in \cite{supply}. The small number of symmetry-allowed parameters in $D_{2d}$, compared with NdAlSi \cite{yang2025crystalline,bauer2009magnetism}, makes the fit considerably better constrained. The starting value of $B_2^0$ was obtained from \cite{wang1971crystal}

\begin{equation}
\theta^a_\mathrm{W}- \theta^c_\mathrm{W} = \frac{3}{10}B^0_2(2J-1)(2J+3)    
\end{equation}

From here, we obtain $B_2^0 = -0.72 $ K, and it is used as the initial starting parameter for the fit. The optimised CEF parameters obtained from the fit are listed in Table~\ref{tab:CEFparams}, in good agreement with the value of $B_2^0$ obtained from Eq. 5. Furthermore, Table~\ref{tab:eigen} provides the full set of obtained energy eigenvalues and their corresponding eigenvectors extracted from the fitted CEF parameters. The GS is found to be a well-isolated doublet at $0$ K, with excited doublets at $16.16$, $32.96$, $63.24$, and $83.9$ K, with a good agreement to the $C_{\mathrm{m}}(T)$ data (Fig.\ref{fig4}c). The $S_{\mathrm{m}}(T)$ computed from the obtained parameters using the expression $S_{m} (T) = \int\frac{C_{m}}{T}dT$ (inset of Fig.~\ref{fig4}c) perfectly describes the experimental data above $T_{\mathrm{C}}$, as below it, the deviation arises because any entropy contribution due to the onset of long-range FM order is not captured by the single-ion CEF Hamiltonian.

\vspace{2mm}

As an independent validation, Fig.~\ref{fig4}d compares the $M(B)$ data at $15$ K (above $T_{\mathrm{C}}$) with the CEF calculation using the same parameters. The isothermal magnetization is computed from the CEF eigenstates as

\begin{equation}
M_i = \frac{g_J \mu_\mathrm{B}}{Z} \sum_n \langle \Psi_n | J_i | \Psi_n \rangle e^{-\beta E_n},
\end{equation}

where $i = x, y, z$, $\beta = 1/k_B T$, and $Z = \sum_n e^{-\beta E_n}$ is the partition function. The eigenvalues $E_n$ and eigenfunctions $|\Psi_n\rangle$ are obtained by diagonalizing the total Hamiltonian including both the CEF and the Zeeman term, with a mean‑field molecular field $\lambda_i M_i$, and represented as \cite{ram2023magnetic}:

\begin{equation}
\mathcal{H} = \mathcal{H}_{\mathrm{CEF}} - g_J \mu_\mathrm{B} J_i \bigl(B + \lambda_i M_i \bigr).
\end{equation}

The good agreement between the calculated curve and the experimental data, without any further adjustment, further supports the derived CEF scheme. The resulting CEF energy level scheme is depicted in Fig.~\ref{fig4}e. Additionally, to verify reproducibility, we performed an identical fit on a second grown crystal from the batch (Sample 2). The extracted CEF parameters agree well with those obtained for Sample 1 (see SI \cite{supply}), confirming the robustness of our analysis and the reliability of the derived CEF scheme.

\vspace{3mm}

\begin{figure}[t]
\centering
\includegraphics[width=0.5\textwidth]{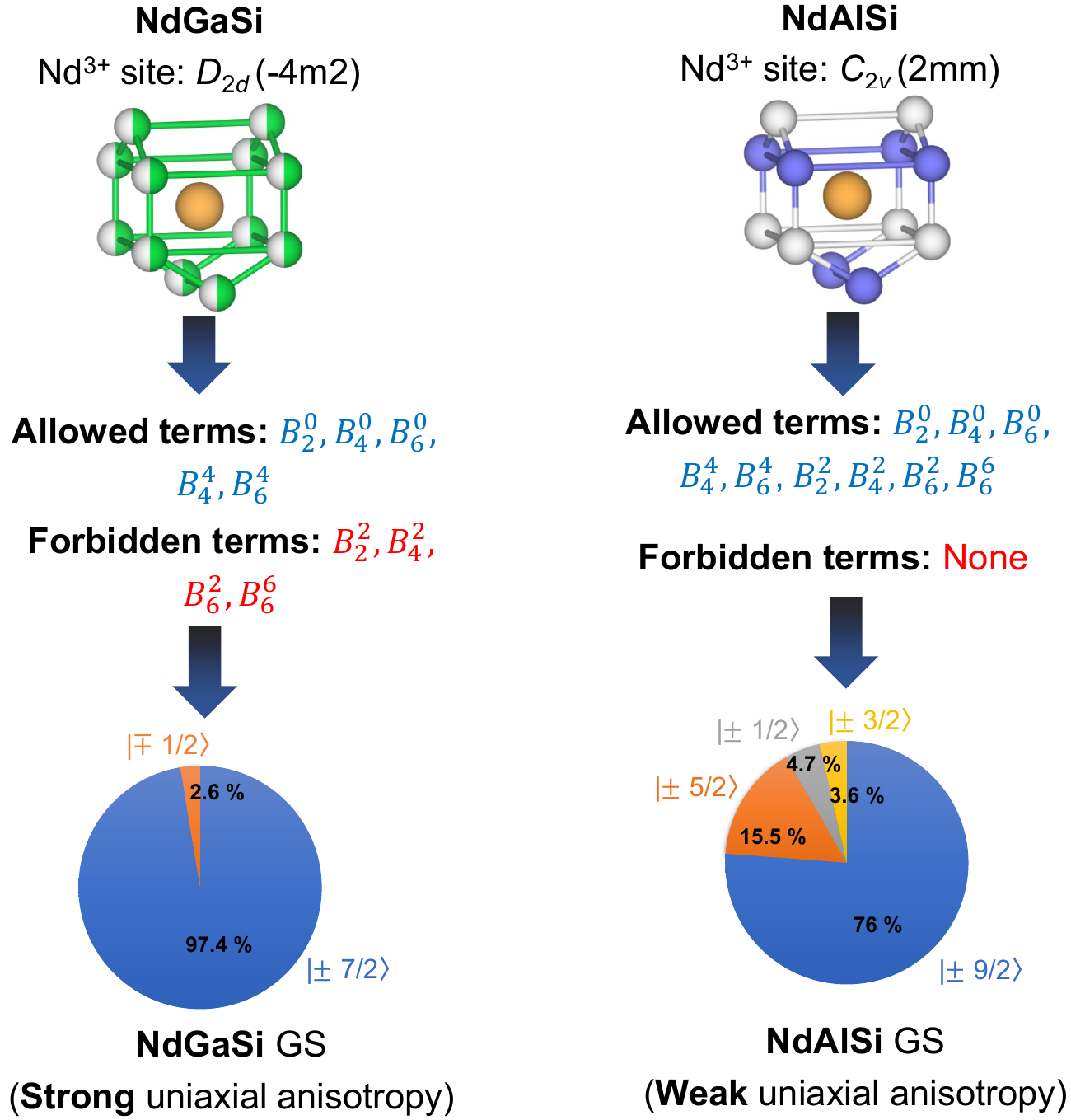}
\caption{An overall schematic representing the pictorial view of the nearest neighbour environment of Nd (orange) for both NdAlSi and NdGaSi, along with the allowed CEF terms, and pie charts representing the weight of various $ |\mathrm{M}\rangle$ states in their respective GS eigenfunctions.   }
\label{fig5}
\end{figure}

\vspace{3mm}
\subsection{Discussion}

The CEF analysis of NdGaSi results in a GS doublet that is ${\sim}97\%$ $|\pm7/2\rangle$ with a small $|\mp1/2\rangle$ admixture (Table~\ref{tab:eigen}). This is the only populated level well below $T_\mathrm{C}$, as observed from the magnetic entropy in the inset of Fig.\ref{fig4}c, and underlies the large uniaxial anisotropy in $\chi(T)$ (Fig.~S2 of the SI \cite{supply}). To understand how this, and the contrast with the INC FIM of NdAlSi \cite{bouaziz2024origin,gaudet2021weyl}, follows from the Nd point symmetry, we compare the CEF Hamiltonians. At the $C_{2v}$ site of NdAlSi \cite{gaudet2021weyl} it contains nine independent parameters:
\begin{multline}
  \mathcal{H}_{\mathrm{CEF}}^{C_{2v}} = B_2^0 O_2^0 + B_2^2 O_2^2
  + B_4^0 O_4^0 + B_4^2 O_4^2 + B_4^4 O_4^4 \\
  + B_6^0 O_6^0 + B_6^2 O_6^2 + B_6^4 O_6^4 + B_6^6 O_6^6
\end{multline}
INS on polycrystalline NdAlSi \cite{yang2025crystalline} determined all nine, with substantial $B_2^2$ and $B_4^2$, showing that the $m=\pm2$ operators are active. At the $D_{2d}$ site of NdGaSi \cite{darone2013rare,saraswati2026accessing}, the Hamiltonian reduces to the five terms (Fig.~\ref{fig5}), and our fit (Table~\ref{tab:CEFparams}) confirms that the effective CEF respects the average $D_{2d}$ symmetry of the disordered lattice. The parameters in Table~\ref{tab:CEFcomp} reflect this difference directly. The axial term $B_2^0$ is nearly twice as large in NdGaSi ($-0.775$~K) as in NdAlSi ($-0.454$~K). The $m=\pm2$ terms $B_2^2$ ($0.363$~K), $B_4^2$ ($-0.0403$~K) and $B_6^2$ ($-0.00155$~K) are significant in NdAlSi but forbidden in NdGaSi. The allowed $B_4^4$ is larger in magnitude in NdGaSi ($0.058$~K vs $-0.0241$~K), yet it only admits a small $|\mp1/2\rangle$ component into the $|\pm7/2\rangle$ doublet.

\vspace{2mm}

\begin{figure}[t]
\centering
\includegraphics[width=0.4\textwidth]{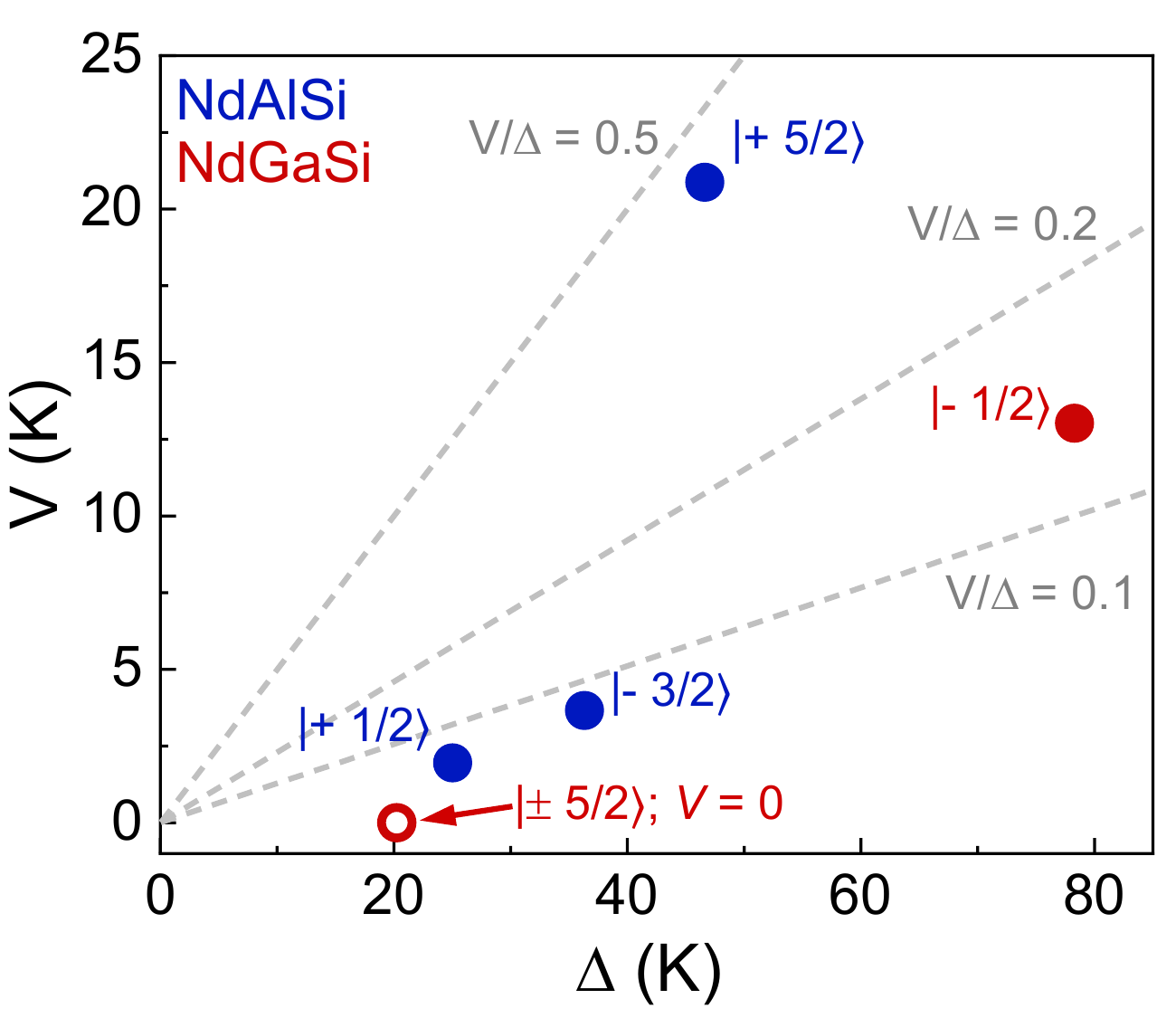}
\caption{Direct coupling of the excited $|M\rangle$ states to the CEF GS of NdAlSi (blue) and NdGaSi (red). The coupling $V = |\langle M|\mathcal{H}_\mathrm{CEF}|M_0\rangle|$ is plotted against the energy gap $\Delta = E_M - E_{M_0}$ between the unperturbed $|M\rangle$ and ground $|M_0\rangle$ states, for $M_0 = 9/2$ (NdAlSi) and $7/2$ (NdGaSi). Dashed lines mark constant first-order admixture amplitudes $|V|/\Delta = 0.1$, $0.2$ and $0.5$. The open circle marks the $|\pm5/2\rangle$ state of NdGaSi ($\Delta\approx20$~K), which has $V=0$ because $D_{2d}$ symmetry forbids the $m=\pm2$ operators}
\label{fig6}
\end{figure}

To understand this, we examine how the Stevens operators mix the free‑ion $|M\rangle$ basis states.  The eigenstates of $\mathcal{H}_{\text{CEF}}$ are superpositions of the $|M\rangle$ states.  To leading order in perturbation theory, the amplitude with which a basis state $|M\rangle$ is mixed into the GS (whose dominant component is $|M_0\rangle = |\pm 7/2\rangle$) is
\begin{equation}
\frac{V}{\Delta} =  \frac{|\langle M|\mathcal{H}_{\text{CEF}}|M_0\rangle|}
       {E^{(0)}_{M} - E^{(0)}_{M_0}},
\end{equation}
where the numerator (\textit{V}) is the off‑diagonal matrix element and the denominator $\Delta$ is the difference between the unperturbed energies of the two basis states.  The diagonal operators $O_l^0$ $(m = 0)$ are responsible for the unperturbed energies $E^{(0)}_{M} = \langle M|\sum_l B_l^0 O_l^0|M\rangle$, whereas the off‑diagonal operators $(m \neq 0)$ provide the matrix elements that couple different $|M\rangle$ states.  The size of each admixture is therefore set by the off-diagonal matrix element and the energy denominator, rather than by the magnitude of the individual $B_l^m$. For NdGaSi this estimate (13~K/78~K $\approx0.17$) reproduces the exact admixture ($0.163$, Table~\ref{tab:eigen}); for NdAlSi (21~K/47~K $\approx0.45$, exact $0.39$ \cite{yang2025crystalline}) it is semi-quantitative. The overall coupling scheme is shown in Fig.~\ref{fig6}, where the coupling $V$ of each excited $|M\rangle$ state to the GS is plotted against its energy gap $\Delta$: in NdAlSi, $|\pm5/2\rangle$ couples strongly across a small gap and lies close to the $|V|/\Delta=0.5$ line, whereas in NdGaSi the only coupled state, $|\mp1/2\rangle$, lies far away in energy, and the nearest excited state, $|\pm5/2\rangle$ ($\Delta\approx20$~K), does not couple at all, because $D_{2d}$ symmetry forbids the $m=\pm2$ operators that would connect it to $|\pm7/2\rangle$.

The $m=\pm2$ operators ($O_2^2$, $O_4^2$, $O_6^2$) change $M$ by $\pm2$ and create linear combinations across the $J=9/2$ manifold: $|\pm9/2\rangle\leftrightarrow|\pm5/2\rangle\leftrightarrow
|\pm1/2\rangle\leftrightarrow|\mp3/2\rangle\leftrightarrow
|\mp7/2\rangle$.  In NdAlSi, the direct coupling $\langle5/2|\mathcal{H}_{\mathrm{CEF}}|9/2\rangle\approx-21$~K, dominated by $B_4^2$ ($\approx-15$~K) and $B_6^2$ ($\approx-8$~K) with only $\approx2$~K from $B_2^2$, acts across an unperturbed separation of only $\approx47$~K. This gives an admixture amplitude of $\approx0.4$, and the network yields the highly entangled GS doublet~\cite{yang2025crystalline} $|\Psi_{\mathrm{0}}^{\mathrm{NdAlSi}}\rangle\approx \mp~0.010|\mp7/2\rangle~\mp~0.190| \mp3/2\rangle\pm~0.217~|\pm1/2\rangle \mp~0.394~|\pm5/2\rangle\mp~0.873~|\pm9/2\rangle$, in which the $|\pm9/2\rangle$ component carries only $76\%$ of the total weight.  By contrast, the sole off‑diagonal operator allowed in $D_{2d}$, $O_4^4=\frac12(J_+^4+J_-^4)$, couples only isolated pairs (e.g.\ $|\pm7/2\rangle\leftrightarrow|\mp1/2\rangle$) and provides no direct path to $|\pm5/2\rangle$ or $|\pm3/2\rangle$.  Moreover, in NdGaSi, $B_2^0$ and $B_4^0$ create a large energy separation between the $|\pm7/2\rangle$ GS and the $|\mp1/2\rangle$ state it couples to, further suppressing the mixing.  The GS of NdGaSi, therefore, remains $\sim$ $97\%$ pure $|\pm7/2\rangle$.  Thus, the connectivity of the mixing network, combined with the energy denominators, is the decisive factor, not the magnitude of individual $B_l^m$ alone. This purity manifests as strong Ising anisotropy, where the moments are strongly confined to the $c$-axis. In contrast, the presence of $B_2^2$ and $B_4^2$ in $C_{2v}$ symmetry introduces transverse mixing, allowing the moments to tilt or cant, which can accommodate the INC helical order observed in NdAlSi.

\vspace{2mm}

For the GS doublet $|\Psi_0^\pm\rangle$, the moments along and perpendicular to $c$ are $m_c=g_J\langle\Psi_0^+|J_z|\Psi_0^+\rangle\,\mu_\mathrm{B}$ and $m_{a,b}=g_J|\langle\Psi_0^+|J_{x,y}|\Psi_0^-\rangle|\,\mu_\mathrm{B}$, with $g_J=8/11$ and $g_{c,ab}=2m_{c,ab}/\mu_\mathrm{B}$.  In NdGaSi, the nearly pure $|\pm7/2\rangle$ GS yields $\langle J_z\rangle\approx3.38$, giving a saturated moment $m_c\approx2.46\,\mu_B$. It is slightly smaller than the observed saturation moment ($\sim ~2.9 ~\mu_\mathrm{B}$). This discrepancy is a known limitation of the point‑charge model, which treats the CEF as a purely electrostatic potential and neglects the inter-ionic exchange interactions \cite{ram2023magnetic,ram2024crystalline,sanki2022valence}. However, the CEF approach remains a powerful and widely adopted framework for rare‑earth intermetallics precisely because, with only a handful of symmetry‑allowed parameters, it simultaneously reproduces the physical properties and the parameters that underpin the magnetic anisotropy. An in-plane moment requires components of $|\Psi_0^+\rangle$ and $|\Psi_0^-\rangle$ that differ by $\Delta M=\pm1$; in NdGaSi these arise only from the small $|\mp1/2\rangle$ and $|\mp9/2\rangle$ admixtures, giving $m_a=m_b=0.13\,\mu_\mathrm{B}$ and $g_c/g_{ab}\approx19$. For NdAlSi, the same expressions give $m_c=2.75\,\mu_\mathrm{B}$, $m_a=0.22\,\mu_\mathrm{B}$ and $m_b=0.49\,\mu_\mathrm{B}$ along the local axes of a single Nd site. The Nd sites occur in two orientations related by the $4_1$ screw axis, which interchanges their local $a$ and $b$ axes, so the bulk in-plane response is tetragonal and given by the average over both. Since the Curie susceptibility scales as $g^2$, we use $g_{ab}=[(g_a^2+g_b^2)/2]^{1/2}\approx0.76$, which gives $g_c/g_{ab}\approx7$ (the single-site ratios are 12.5 and 5.6).  The much larger ratio in NdGaSi is consistent with the Ising-like anisotropy observed with the easy axis along the $c$ direction. The entire comparison is pictorially represented in the form of a schematic in Fig. \ref{fig5}.

\begin{figure}[t]
\centering
\includegraphics[width=0.5\textwidth]{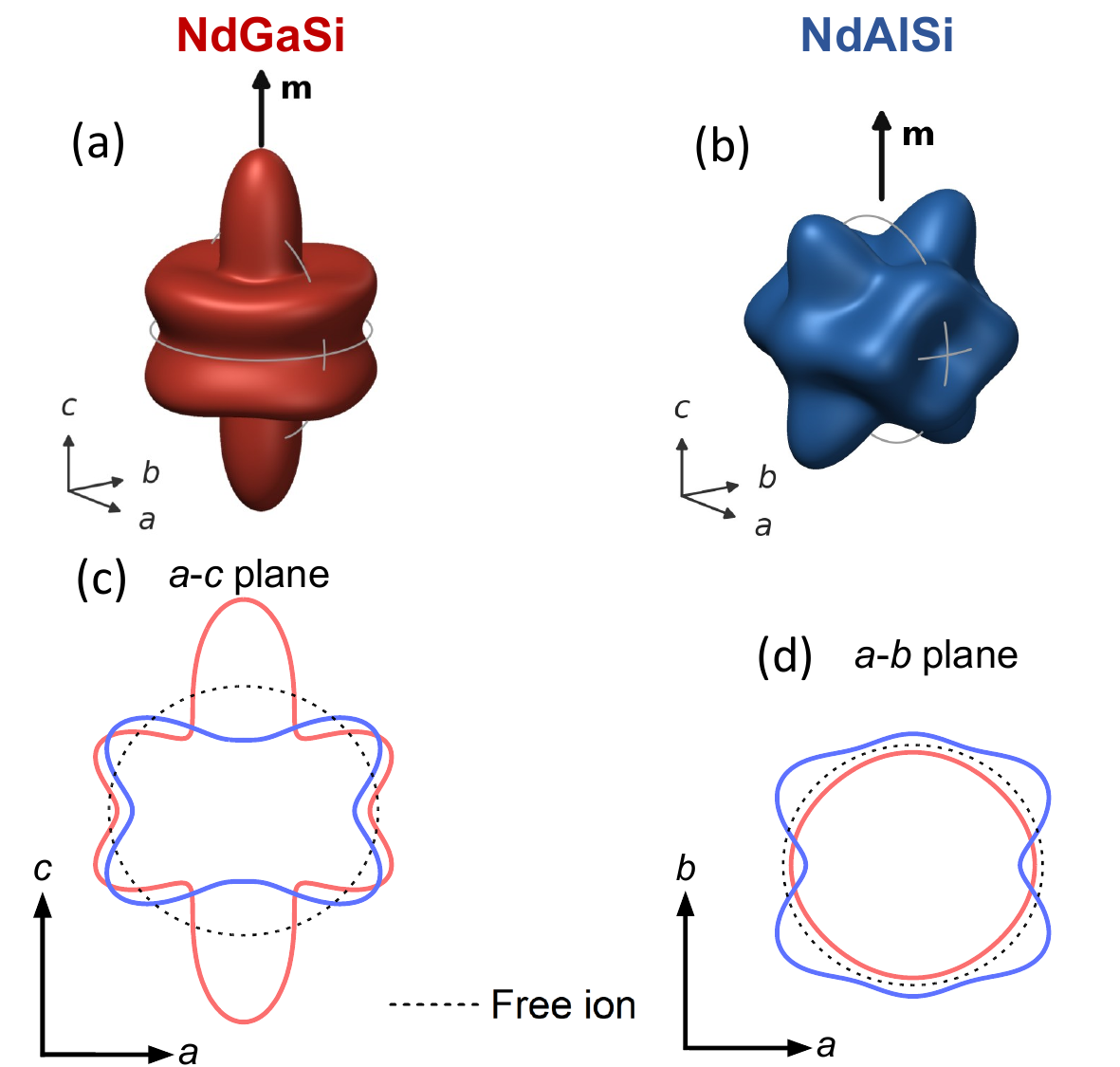}
\caption{Angular $4f$ charge density plot of the Nd$^{3+}$ CEF GS doublet, calculated from Eq.~10 with the CEF parameters of (a) NdGaSi (Table~\ref{tab:eigen}) and (b) NdAlSi (Ref.~\cite{yang2025crystalline}). Grey rings mark the free-ion sphere in the $ab$ and $ac$ planes, and arrows indicate the GS doublet moment $\mathbf{m}$ along $c$. (c) \textit{ac} and (d) \textit{ab} plane cross sections of the charge density of NdGaSi (red) and NdAlSi (blue), compared with the free-ion sphere (dotted).}
\label{fig7}
\end{figure}

\vspace{2mm}

The GS purity also has a direct real-space picture in the shape of the $4f$ charge cloud as represented in Figs. \ref{fig7}a and \ref{fig7}b. Following Sievers' procedure~\cite{sievers1982asphericity}, the angular $4f$ charge density of the GS follows from the Stevens operator equivalence $\sum_i f_l^m(\hat r_i)=\Theta_l O_l^m$~\cite{stevens1952matrix,hutchings1964point}, where $f_l^m$ is the polynomial corresponding to $O_l^m$ (e.g.\ $O_2^0\leftrightarrow3z^2-r^2$):
\begin{equation}
\rho(\theta,\phi)=\frac{3}{4\pi}+\sum_{l,m}\frac{\Theta_l\langle O_l^m\rangle}{\int (f_l^m)^2\,d\Omega}\,f_l^m(\theta,\phi),
\end{equation}
where $\Theta_l$ are the Stevens factors ($\Theta_2=\alpha_J$, $\Theta_4=\beta_J$, $\Theta_6=\gamma_J$) and $\langle O_l^m\rangle$ is taken in the GS doublet (derivation in Sec.~S6 of the SI~\cite{supply}). The nearly pure NdGaSi doublet gives a cloud that is elongated along $c$ and almost axially symmetric about it. The mixed NdAlSi doublet instead gives a flattened, locally two-fold cloud. The cross sections in Figs.~\ref{fig7}c and \ref{fig7}d show this directly. In the $ac$ plane, the NdGaSi cloud is stretched along $c$ while the NdAlSi cloud is squeezed along $c$. Similarly, in the $ab$ plane, the NdGaSi section is nearly circular, while the NdAlSi section is clearly distorted. Because spin--orbit coupling ties the moment to this charge distribution~\cite{rinehart2011exploiting}, the axially symmetric cloud of NdGaSi leaves no preferred in-plane direction and confines the moment to $c$, consistent with the vanishing in-plane moment found by SCND.

\vspace{2mm}

Frustrated intra- and interlayer SX can result in a finite-$\mathbf{q}$ modulation, but whether such a modulation is realised also depends on how easily the ordered moment can change from site to site, either in direction (canting) or in length (amplitde modulation). In NdGaSi, the nearly pure $|\pm7/2\rangle$ GS doublet carries almost no in-plane moment ($m_a=m_b=0.13\,\mu_\mathrm{B}$). Any moment component away from $c$ must therefore be induced by mixing in excited doublets, the lowest of which lies 16~K above the GS, and at low temperature the moment length along $c$ is likewise fixed by the GS. Both canting and modulation thus cost CEF energy, for which the system prefers a uniform collinear FM state. In NdAlSi, by contrast, the mixed GS doublet already carries an in-plane moment ($m_a=0.22$ and $m_b=0.49\,\mu_\mathrm{B}$), so canted and modulated structures can form without this cost, consistent with the magnetic order observed~\cite{gaudet2021weyl}. EXAFS identifies the structural origin of this difference: the ordered Al/Si sublattice lowers the Nd site symmetry to $C_{2v}$ and allows the $m=\pm2$ mixing that produces the mixed GS of NdAlSi.

\vspace{2mm}

In the \textit{R}AlSi semimetals, the magnetic order is driven primarily by frustrated short-range SX rather than by Weyl-mediated RKKY or DMI \cite{bouaziz2024origin}. In noncentrosymmetric NdAlSi, the DMI, $\mathcal{H}_{\mathrm{DM}}=\sum_{\langle i,j\rangle}\mathbf{D}_{ij}\cdot(\mathbf{S}_i\times\mathbf{S}_j)$, is allowed, but first-principles calculations show that it is at least an order of magnitude smaller than the frustrated isotropic SX, and switching it off leaves the finite-$\mathbf{q}$ peak of $J(\mathbf{q})$ essentially unchanged: DMI fixes the chirality of the helix, not its existence or period. In centrosymmetric NdGaSi, the uniform DMI allowed by the polar structure of NdAlSi is absent. ($\mathbf{D}_{ij}=0$).

\vspace{2mm}

The different GS of the two compounds therefore cannot be attributed to DMI, but reflect how the CEF-determined moment responds to frustrated SX: a rigid $|\pm7/2\rangle$ doublet resists modulation, whereas the mixed $C_{2v}$ doublet can. The structural origin is the Ga/Si antisite disorder, which controls the CEF GS purity. Whether the exchange in NdGaSi retains the same finite-$\mathbf{q}$ tendency is not resolved by the present experiments; what is established is that its rigid Ising doublet strongly suppresses the channel through which such a tendency could be realised.

\begin{table}[t]
\centering
\caption{Comparison of the CEF parameters for NdGaSi (this work) and NdAlSi 
(from Ref.~\cite{yang2025crystalline}). All values are given in kelvin. 
}
\label{tab:CEFcomp}
\renewcommand{\arraystretch}{1.2}
\begin{ruledtabular}
\begin{tabular}{l c c}
Parameter (K) & NdGaSi   & NdAlSi  \\
\hline
$B_2^0$   & $-0.775$     & $-0.454$ \\
$B_2^2$   & forbidden    & $\;\;0.363$ \\
$B_4^0$   & $\;\;0.015$  & $-0.0089$ \\
$B_4^2$   & forbidden    & $-0.0403$ \\
$B_4^4$   & $\;\;0.058$  & $-0.0241$ \\
$B_6^0$   & $\;\;0^*$     & $\;\;3.95\times10^{-5}$ \\
$B_6^2$   & forbidden    & $-0.00155$ \\
$B_6^4$   & $\;\;0^*$     & $\;\;7.71\times10^{-4}$ \\
$B_6^6$   & forbidden    & $-0.00111$ \\
\end{tabular}
\end{ruledtabular}
\vspace{-2mm}
\footnotesize * Sixth‑order terms were fixed to zero in the fit.\\
\end{table}

\section{CONCLUSION}

\vspace{3mm}

In summary, we have combined single-crystal growth, magnetization, specific heat, neutron diffraction and XAS with a CEF analysis to establish the origin of the large anisotropy and collinear FM order in NdGaSi. Ga/Si antisite disorder, confirmed by neutron diffraction and EXAFS, stabilises the centrosymmetric $I4_1/amd$ structure and places Nd$^{3+}$ at a $D_{2d}$ site, forbidding all $m=\pm2$ Stevens operators. The GS therefore remains a $\sim97\%$ pure $|\pm7/2\rangle$ Kramers doublet with strong Ising anisotropy along $c$ ($g_c/g_{ab}\approx19$, compared with $\approx7$ in NdAlSi), which accounts for the large moment along \textit{c}, and the sharp FOMT at $7.9$~T for $B\perp c$. In NdAlSi, the ordered Al/Si arrangement lowers the Nd symmetry to $C_{2v}$ and produces a highly entangled CEF GS. The decisive mechanism is not a direct competition between anisotropy and exchange, but whether the CEF GS purity leaves the ordered moment rigid (NdGaSi) or able to respond to a finite-$\mathbf{q}$ exchange field (NdAlSi). 

\vspace{2mm}

Disorder on the non-magnetic sites is often observed in rare-earth intermetallics, being usually regarded as a structural feature with little emphasis on the magnetism. However, our study on NdGaSi and NdAlSi shows that although Al and Ga are isoelectronic, whether the non-magnetic atoms are ordered or randomly mixed sets the symmetry of the Nd site and, through it, the nature of the CEF, ultimately leading to the preference of the magnetic GS. Antisite order/disorder on the $p$-block sublattice is therefore not merely a structural detail but a parameter for controlling single-ion anisotropy and selecting between a frustrated INC FIM and a robust collinear FM. This principle may apply to other rare-earth families, where deliberate control of site mixing could tune the CEF GS purity and, consequently, the magnetic and topological properties.

\vspace{3mm}

\section*{ACKNOWLEDGEMENTS}
\vspace{3mm}
NK acknowledges DST for financial support through Grant Sanction No. CRG/2021/002747 and the Max Planck Society for funding under the Max Planck-India partner group project. This research project made use of the instrumentation facility provided by the Technical Research Centre (TRC) at the S. N. Bose National Centre for Basic Sciences, under the Department of Science and Technology, Government of India. JS\ acknowledges the Anusandhan National Research Foundation (ANRF), Government of India, for the Ramanujan Fellowship (Grant No.\ RJF/2019/000046 (SQUID-1986-JS3632)) and POWER Grant No.\ SPG/2021/003136. SN acknowledges DESY (Hamburg, Germany), a member of the Helmholtz Association HGF, for the provision of experimental facilities. Parts of this research were carried out at PETRA III, P64, and SN would like to thank Dr Aleksandr Kalinko for beamline guidance. 

\bibliography{reference}

\end{document}